\documentclass{emulateapj}

\def\bra#1{\left[ #1 \right]}

\shorttitle{GRB~060206 Early-Time Light Curves: GRB Energetics and Environment}
\shortauthors{Monfardini et al.}

\begin{document}


\title{High Quality Early Time Light Curves of GRB~060206: Implications for Gamma
 Ray Burst Environments and Energetics}

\author{A.~Monfardini\altaffilmark{1}, S.~Kobayashi, C.~Guidorzi, D.~Carter, C.~G.~Mundell, \\D.~F.~Bersier,
A.~Gomboc\altaffilmark{2}, A.~Melandri, C.~J.~Mottram, R.~J.~Smith, I.~A.~Steele}
\affil{Astrophysics Research Institute, Liverpool John Moores University,
Twelve Quays House, Birkenhead, CH41 1LD, UK}

\email{(am, sk, crg, dxc, cgm, dfb, ag, axm, cjm, rjs, ias)@astro.livjm.ac.uk}

\altaffiltext{1}{Also: ITC--IRST and INFN, Trento, via Sommarive, 18 38050 Povo (TN), Italy.}
\altaffiltext{2}{Present address: FMF, University in Ljubljana, Jadranska 19, 1000 Ljubljana, Slovenia.}

\begin{abstract} 

The 2-m robotic Liverpool Telescope ({\em LT}) reacted promptly to the high-redshift ($z=4.048$) 
gamma--ray burst GRB~060206.
The afterglow was identified automatically and
multicolor $r'i'z'$ imaging program was triggered without human intervention. 
Combining our data with those obtained from later follow-ups
provides a well-sampled 
optical light-curve from 5~minutes to $>$2d after the gamma event.     
The light-curve is highly structured with at least three bumps evident in the first 75 minutes, including a major rebrightening 
($\Delta r'\approx-1.6$ at $t\approx3000s$),
interpreted as late energy injection.
At early time ($t\approx440s$), we find evidence for fast ($\Delta t_{rest}<4s<<t$)
variability, indicating on-going internal-engine activity.
We emphasise that a low redshift GRB ($z<1$) with similar 
intrinsic properties would have been interpreted completely differently
due to undersampling of the light curve in the rest frame at early times; the
light-curve behaviour of GRB~060206 should therefore not be considered peculiar.
Finally, although the observed late-time steepening of the
optical light curve resembles a jet break if taken in isolation, the
lack of a corresponding change in the X-ray slope rules out a
jet-break interpretation. Traditionally, GRB jet breaks have been
inferred from optical data in the absence of simultaneous X-ray data.
We suggest therefore that current estimates of the jet opening angle
distribution might be biased by events like GRB~060206. Consequently,
the GRB explosion energy distribution and event rates may have to be
revised.

\end{abstract}

\keywords{gamma rays: bursts - cosmology: observations}

\section{Introduction} The {\em Swift} satellite~\citep{Gehrels04},
launched in November 2004 and specifically designed to study Gamma Ray
Bursts, is providing an overall rate of about 100 real-time gamma
localizations per year. A considerable fraction ($\approx75\%$) are
reobserved within minutes by the on-board X-ray ({\em XRT}) and
UV-optical ({\em UVOT}) narrow field-of-view instruments, whilst an
increasing number of ground-based robotic facilities are participating
actively in the worldwide follow-up program by responding rapidly to
satellite alerts.  As a result, the number of GRBs with early-time
light curves and spectroscopically determined redshift has increased
dramatically over the last 18 months. Thus, in addition to important
recent results such as the detection of short burst afterglows and the
observation of the most distant (z$>$6) GRB observed to
date~\citep{Haislip05,Antonelli05}), the following facts are emerging:
{\it a)} a higher mean redshift ($z\approx3$) for Swift GRBs
~\citep{Jakobsson06} compared with those detected in the pre-Swift era
($z\approx1$~\citep{Berger05});
{\it b)}  990123-like \citep{Akerlof99}
prompt flashes are on average weaker and less frequent than expected;
{\it c)} a population of faint bursts (truly {\it
dark} only in absence of an early, deep follow-up) providing useful
constraints on existing optical suppression
models~\citep{Roming05,Oates06}; {\it d)} huge, rapidly rising and
decaying flares observed in the X-ray light-curves, indicating
probable late-time activity of the inner engine~\citep{Burrows05}.
The growing evidence for continued activity of the central engine, the
detailed physics of which remains elusive, and the opportunity to
probe the interaction of expanding shells with the
circumburst/interstellar medium, particularly at high redshift, demand
dedicated prompt followup observing programs. Large robotic telescopes
such as the 2-m Liverpool Telescope are optimised for such followup and are proving effective~\citep{Guidorzi05, Guidorzi06}.

Early afterglow light curves exhibit a considerable variety and rebrightenings
(hereafter called {\it bumps}) are becoming more frequently
observed. GRB~021004 \citep{Bersier03,Matheson03,Holland03} remains one of the best studied events
($z=2.335$) and  a number of different models have been invoked to explain its temporal behaviour: variable density
profile~\citep{Lazzati02}, passage of the break
frequency~\citep{kobayashi03}, refreshed shocks
\citep{Nakar03,Bjornsson04,deUgarte05} , angular dependence of the
energy profile on the jet structure \citep{Nakar03}.  Another
interesting case is GRB~030329 ($z=0.17$), preferably modeled
according to the refreshed shocks (late energy-injections) scenario
\citep{Huang06}.  In this case, initially slower shells catch up with the
external shock front, which, in the meantime, is decelerating due to the
interaction with the surrounding medium. The early afterglow of  GRB~050502a ($z=3.793$)
exhibited a  smooth decay, well-fitted by a power-law
with index $\alpha=1.2$ ($F_{\nu}\propto t^{-\alpha}$), followed by a
 an achromatic bump emerging at $t_{rest}\approx600s$ that has been
 interpreted by \citet{Guidorzi05} as due to density clumps in a
 uniform medium.  More recently, an increasing number of events, in
 particular at high-z, exhibit rebrightenings \citep{Rykoff06,
 Stanek06}. These observations support the hypothesis such behaviour
 might not be unusual.  We report here multicolor robotic
 observations of GRB~060206 carried out by the Liverpool Telescope
 ({\em LT}) and revealing a structured behaviour of the early
 lightcurve. The 2-m aperture of the {\em LT}, the optimal observing
 conditions and a precise SDSS pre-burst field calibration
 ~\citep{Cool06} contributed to the remarkable quality of the dataset.
 Timely later photometry from {\em RAPTOR} \citep{Wozniak06} and {\em
 MDM} \citep{Stanek06} allowed us to compile an exceptionally
 information-rich light-curve catching the different phases of the
 early afterglow.

\section{Observations and Analysis}

On 2006 Feb 06, 04:46:53 UT\footnote{t=0 throughout the paper.} {\it Swift-BAT} 
detected GRB~060206 (trigger 180455) at Galactic Coordinates l=78.07deg, b=78.28deg\footnote{According to
\citet{Schlegel98} $E(B-V)=0.013$. The estimated extinction, from \citet{Cardelli89} curve, 
is  $A_{r'}=0.036$, $A_{i'}=0.027$ and $A_{z'}=0.019$.} (3 arc-min 90\% containment radius).
In the $15-350keV$ band the prompt gamma profile is well described by a single quasi-gaussian peak with duration $7s\pm2s$.
{\em Swift-XRT} began observing 65s after the BAT trigger time and
entered the South Atlantic Anomaly soon afterwards \citep{Morris06}.
UVOT started the on-target monitoring practically at the same time.
As usual, a number of ground-based facilities were activated to follow-up the event. 
The redshift has been spectroscopically measured to be $z=4.048$ \citep{Fynbo06}.
Robotically triggered photometric observations 
with the  Liverpool Telescope began at $t=309s$ after the GRB, at an airmass of $1.01$. 
A sequence of $2\times3\times10s$ $r'$ exposures, named {\it detection mode}, aims at finding potential afterglow 
candidates and automatically triggers the best suitable early follow-up observing program. Details of the 
procedure are reported elsewhere by \citet{Guidorzi06}. In the GRB~060206 case, a repeated sequence of $r'i'z'$ 
120s exposures was initiated.
The afterglow is detected with high S/N in all the single exposures. 
Calibration was facilitated and greatly improved by the availability of SDSS pre-burst photometry
\citep{Cool06}.
We adopted 5 calibrated star-like PSF field objects to adjust the zero-point of the single images.
The zero points were stable throughout the observation
sequence, and consistent with nominal LT values\footnote{Confirmed also by the CMT in La Palma. 
Nightly averaged atmospheric extinction: r'=0.076. http://www.ast.cam.ac.uk/$\sim$dwe/SRF/camc\_extinction.html}. 
Photometry was carried out independently using the Starlink/GAIA and DoPHOT programs.
The datasets are in complete agreement with each other, and the results 
have been combined to further reduce the systematics. 
Magnitudes are converted into flux densities $F_{\nu}$ (mJy) following \citet{Fukugita96}.
Results are summarized in Table~\ref{tab:obs}.

\placetable{tab:obs}

\begin{figure}
\epsscale{.80}
\plotone{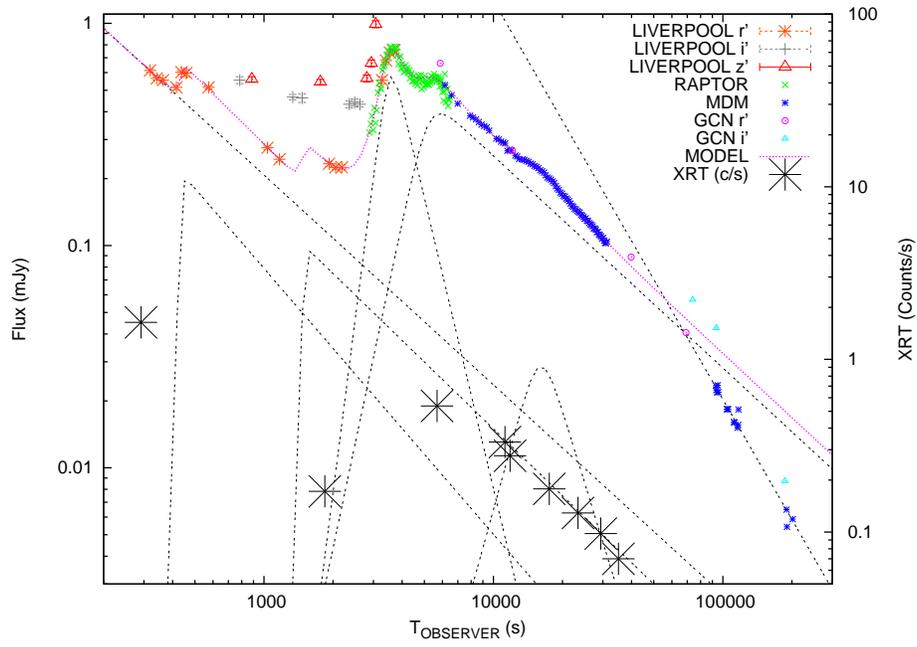}
\caption{Complete GRB~060206 light-curve. Early ($t<3600s$) multi--color $r'i'z'$ photometry from Liverpool Telescope
robotic observations. Later data from \citet{Wozniak06}, \citet{Stanek06}, \citet{Malesani06}, \citet{Reichart06}, 
\citet{Haislip06}. Also shown: analytical model best fit, separate contributions (dotted lines), {\em XRT} 
counts rates (right-side axis).\label{fig:LC}}
\end{figure}

Figure~\ref{fig:LC} shows the complete lightcurve
of GRB~060206. 100 {\em RAPTOR} \citep{Wozniak06} and 
101 {\em MDM} \citep{Stanek06} photometric points have been included together with selected GCNs.
The RAPTOR unfiltered values, calibrated with respect to USNOB1.0-R2 magnitudes \citep{Wozniak06}, 
have been flux-recalibrated with respect to the earlier {\em LT} data. This is 
equivalent of shifting the {\em RAPTOR} data by +0.3 magnitudes. This approach relies on a relatively stable
colour. At a first glance this assumption might look uncertain since we are actually 
presenting evidence of a spectral evolution across the bump. However, by assuming a realistic 
$\Delta (r'-i')_{bump} \approx 0.4$, a rather small $\Delta r'_{sys}\approx-0.07$ is estimated. 
This systematic correction has been added to the synthetic $r'$ late photometry data. 
In any case this isn't in any way affecting our conclusions.
The {\em MDM} data, as in \citet{Stanek06}, have been shifted by +0.22 magnitudes with respect to {\em RAPTOR} 
and accordingly converted into physical units. 
{\em XRT} data have been processed with the ftool 'xrtpipeline'
 to produce cleaned level II event files,
applying default screening constraints. We select only 
events with grade 0 for both {\em PC} (Photon Counting)
and {\em WT} (Windowed Timing) modes. For {\em PC} mode, we 
initially considered a 20 pixel-radius circle centred on the source.
From the PSF profile, it was clear the data were partially affected by
pile-up; following \citet{Vaughan06}, we corrected for this
by excluding photons within a 5-pixel radius inner circle
and then renormalised the resulting light curve for the
fraction of the PSF profile considered.
Background spectrum and light curve have been taken from
four 50-pixel radius circles with no sources.
No significant pile-up was found to affect WT data, following
the procedure described in \citet{Romano06}.
We adopted empirical ancillary files for the spectra
created with the ftool 'xrtmkarf'. We used the latest
spectral response matrices in the Calibration Database
(CALDB 2.3).

\placefigure{fig:LC}

\subsection{r' Lightcurve Fit}

The analytical model adopted for the fit, up to the break at
$\sim3\times10^4~s$, is:

\begin{equation}
F_{\nu}(t)=\sum_j F_{j}\cdot\sqrt[n]{\frac{2}{(t/t_j)^{-\alpha_{1_{j}}\cdot~n}+(t/t_j)^{-\alpha_{2_{j}}\cdot~n}}}
\end{equation}

This is the sum of a number of smoothly connected broken power-laws
\citep{Beuermann99}. $\alpha_1$ and $\alpha_2$ depend on j and are the
exponents at $t<<t_j$ and $t>>t_j$ and $t_j$ the bump (break)
time. When $\alpha_1<0$ and $\alpha_2>0$, a bump is produced. A
typical light-curve break, on the other hand, is obtained when
$\alpha_1,\alpha_2 > 0$ and $\alpha_2\neq\alpha_1$.  The constant $n$
controls the smoothness of the transitions, but has a relatively small
influence in this context (see Table~\ref{tab:fittime}).  $F_j$ is the
individual flux contribution at the bump (break) time. In the case of
GRB~060206 the $j$ index runs from 0 to 5 (see Figure~\ref{fig:LC}). The $j=0$ segment is, by
definition, associated with $t_j=0$ and is therefore a simple
power-law fitting the earliest {\em LT} photometric points.  The $j=1$
segment accounts for the sharp, earliest bump occurring at
$t\approx440s$ and the subsequent decay. The next segment describes
three r' points lying at $t\approx2000s$. However, this fit at 2000s
is not unique as one may also fit the flattening of the curve around
this time with $\Delta\alpha\approx-0.8$, beginning immediately after
$t=1200s$.  Although this scenario may seem simpler than introducing a
bump component via $j=2$, assuming an achromatic flattening in the
$r'$ band curve for $t=1200s:1900s$ (constrained by $r'$ sampling)
would in turn imply a rebrightening in $z'$ and $i'$.  This confirms
that a bump is required in this time interval. In our analytical model
the bump time is arbitrarily fixed to $t_2=1500s$, and the shape fixed
by the earliest case template. The only variable is the peak flux
$F_2$ that is in any case non significantly affected by $t_2$.
In other words, a small rebrightening at $t\approx1500s$ is necessary
because it is not possible to fit the light curve in all 3 colors ($r'i'z'$)
without including a bump at that time.
The next segments, $j=3,4$ provide a good fit to the biggest double-peaked
bump ($t_{3}=3529s\pm26s$, $t_{4}=5234s\pm79s$). Most importantly, the
cosmological time dilation $(1+z)$ implies that the intense activity
described so far took place in less than 15 minutes in the rest frame.
The last feature before the final break, possibly associated to a density variation, 
takes place at $t\approx16000s$ and is well described by the $j=5$ term. 
In absence of a systematic photometric sampling close to the late break 
($t\approx0.6d$), we fitted the $t>1d$ {\em MDM} points with a simple, 
independent power-law. 
The analytical model validity is thus limited to $t<50000s$, before the 
occurrence of the inferred break.
The result is a steepening from $\alpha_1=0.95$ ($t<<t_{break}$) to 
$\alpha_2\approx1.8$ ($t>>t_{break}$).
We note that these values differ slightly from those found by \citet{Stanek06} 
who derived $\alpha_1 \sim 0.7$ to $\alpha_2\approx2.0$. The difference is to be attributed 
probably to the $j=5$ bump that is recognized as a separate feature in our view, but
included in the general pre-bump slope in \citet{Stanek06}.
That late break nature is discussed in more detail in the next Section.\\
The analytical model fit was performed separately in three different time ranges:
$300:2500s$ ($6$ Degrees Of Freedom), $2500s:15000s$ ($126$ $DOF$) and
$7000s:50000s$ ($74$ $DOF$). Since the fit parameters are common, the
procedure is iterated to let them converge to the same values in the
three intervals. Convergence is achieved after $\sim10$ iterations,
and the result is then stable indicating that a true global minimum
has been reached. A summary of the parameters resulting from the fit
is reported in Table~\ref{tab:fittime}.  In Figure \ref{fig:LC} the
single components are shown together with the resultant fitted curve.

\placetable{tab:fittime}

\subsection{Multi-epoch Spectral Energy Distribution Analysis}

\begin{figure}
\epsscale{.80}
\plotone{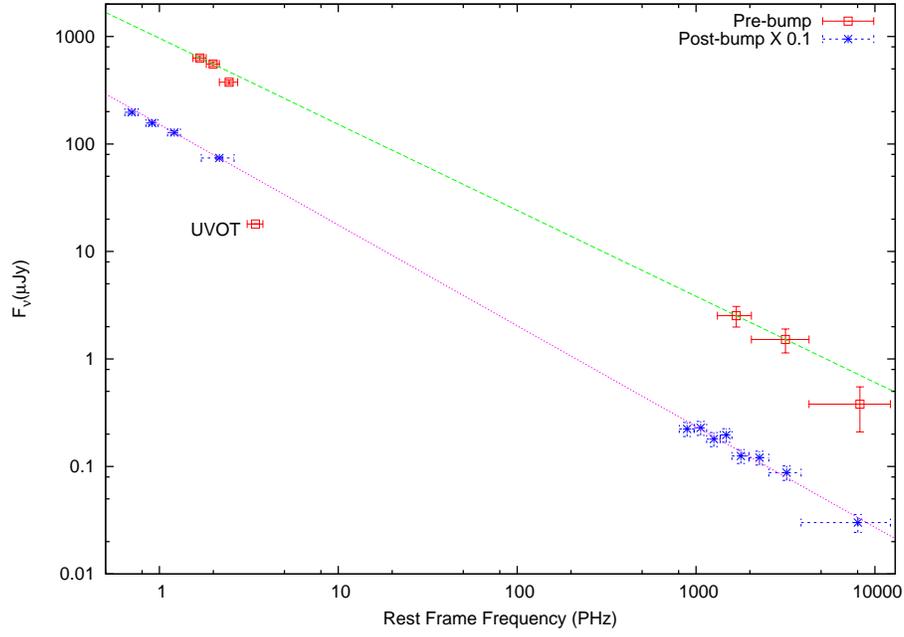}
\caption{GRB~060206 pre- and post-bump Rest--frame SED ($t=785s$ and $t=5680s$). Optical points from Liverpool 
Telescope robotic observations, \citet{Stanek06} and \citet{Boyd06}. The $5680s$ fluxes are divided by 10. 
Dotted lines: $\beta=0.80$ and $\beta=0.93$ power-law fitting respectively pre- and post-bump points.
In the text we report evidence of a significant {\em SED} evolution occurring in $t=1000s:3000s$.
The fit results are summarized in Table~\ref{tab:spectra}.
\label{fig:SED}}
\end{figure}

Figure~\ref{fig:SED} shows the pre-bump rest--frame {\em SED} interpolated at $t=785s$, where the $r'$ lightcurve is relatively 
smooth, from {\em LT} and {\em XRT} data. 
The {\em XRT} spectral shape at this epoch have been assumed coincident 
with the PC early template ($t=141s:341s$). The X-ray flux at $t=785s$, is derived from an $\alpha=1$ (Table \ref{tab:fittime}) power-law extrapolation. 
This seems a quite reasonable assumption 
(see Figure~\ref{fig:LC}). Further, no significant differences are seen between the {\em XRT} spectra extracted from PC and WT modes.
An extrapolated {\em UVOT-B} point has been added to provide important qualitative constraints. 
The average time behaviour of the B light-curve is assumed to follow the well-sampled $r'$ case. 
An $F_{r'}(t)$ integral is performed over the {\em UVOT} time span 
($t=1662s:34946s$, \citet{Boyd06}). The averaged $F_{r'}^{aver}$ is directly compared to $F_{B}^{UVOT}$. 
The following simple ratio relation provides an estimate of $F_{B}$ 
corresponding to the pre-bump epoch of interest:

\begin{equation}
F_{r'}^{aver}:F_{r'}(T)=F_{B}^{UVOT}:F_{B}(T)
\end{equation}

At $t=785s$ the four optical points are not reconcilable with a single power-law. The $r'$ flux is slightly
affected, according to \citet{Fynbo06}, by the deep Damped Lyman Alpha {\em DLA} absorption feature. 
The $i'$ and $z'$ values, on the other hand, can be very well extrapolated up to $\nu_{rest}\approx4840 PHz$
following a rather typical $\beta=0.8$ slope ($F_{\nu}\propto\nu^{-\beta}$). 
The $UVOT-B$ point qualitatively confirms the pronounced $HI$ absorption and is not included in the 
fitting procedure.
Even if sparsely sampled, the positions of the $i'$ and $z'$ photometric points in the time 
interval $t=1000s:3000s$ clearly indicate either a {\em SED} evolution or an additional unresolved
activity (or both combined).
A rigid shift\footnote{The multiplicative factors for $i'$ and $z'$ are determined according to the 
earliest epoch and are respectively $1.50\pm0.05$ and $1.72\pm0.05$.} 
of the $r'$ light-curve, interpreted either with the bump or the flattening, 
indicate that the $i'$ and $z'$ fluxes are significantly higher than expected. According to the photometric 
errors, the disagreement significance has been estimated in $4\sigma$ in the bump scenario ($i'$, $9\sigma$ 
for $z'$). Analogous results assuming an achromatic flattening ($5\sigma$ and $11\sigma$).
We suggest that the {\em SED} evolution scenario is more conservative, but caution that 
a generally accepted quantitative picture would require better sampling of the light curves.
As an additional note, we remark that the simple rescaling described in footnote is again able to account for
the $z'$ points calculated during the big bump rising slope. This might be a possible indication, even if not 
conclusive, that the {\em SED} is already relaxed before the bump reaches its peak.

\placefigure{fig:SED}

In Figure \ref{fig:SED} we show the post-bump SED at $t=5680s$ based on the {\em PAIRITIEL} 
JHK detections \citep{Alatalo06} ($t_{start}=2.91h$) and on the earliest {\em MDM} $R$ 
point\footnote{We adopted the original \citet{Stanek06} $R$ flux 
instead of the estimated $r'$. The $SED$ is thus completely unaffected by the 
LT-to-RAPTOR conversion systematics described in detail in the text.}.
Both are back-extrapolated according to an $\alpha=0.95$ temporal slope (table~\ref{tab:fittime}). 
At this late time, already on the smooth side of the afterglow, the IR-X  fluxes ($\nu_{rest}=0.7:8000 PHz$) 
are well fitted by a common $\beta=0.93\pm0.02$ power-law ($\chi^{2}/d.o.f.=3.46/10$). However, a broken 
power-law with fixed indices $\beta_{OPT}=0.7$, $\beta_{X}=1.2$ is not ruled out 
($\chi^{2}/d.o.f.=10.04/10$). The latter is considered in the light of the interpretative model discussed 
later.\\
Post-break IR photometry \citep{Terada06} and {\em MDM} late observations give $\beta_{IR-O}=0.7\pm0.3$.
A summary of the spectral results is given in Table~\ref{tab:spectra}.

\section{Discussion}
\label{sec:disc}

\subsection{Re-Brightening}
After the remarkable rebrightening at $\sim 1$hr and before the break
around $t\sim 5\times10^4$ s, the optical flux is described
by a single power law with $\alpha \sim 1$, except small wiggles.
This indicates that the blast wave radiating the afterglow had a
significant transition from one Blandford-McKee (BM; Blandford\&McKee 1976)
solution to another BM solution around $t = 1$ hr. Since blast wave
solutions depend only on two parameters (explosion energy and ambient
density) possible scenarios are energy injection (Rees\& M\'esz\'aros
1998; Kumar\& Piran 2000; Sari \& M\'esz\'aros 2000) and a density-jump
medium (Ramirez-Ruiz et al. 2001; Dai \& Lu 2002). The degeneracy is
broken by  the X-ray observations. Since there is no signature of the
cooling break in the decay phase of the X-ray afterglow (Morris et
2006), the X-ray band is likely to be located above the cooling frequency
at $t \sim 1$ hr. This implies a high ambient density, consistent with
spectroscopic determinations \citep{Fynbo06}. The late time break in the optical light curve also
favors $\nu_x > \nu_c$ at $\sim 1$ hr as we will discuss below. If the
X-ray band is above the cooling frequency, the flux does not depend on
the ambient density \citep{Freedman01}. We can conclude that the X-ray rebrightening is not
due to a jump in the ambient density.

According to \citet{Kumar00}, in the absence of a density enhancement, the flux depends on the
explosion energy as: 


\begin{eqnarray}
F_{\nu}\propto E^{(p+3)/4}&~~&{\nu_m<\nu <\nu_c}\\
F_{\nu}\propto E^{(p+2)/4}&~~&{ \nu>\nu_c}
\end{eqnarray}

Since the
optical flux increased by a  factor of $\sim 4$, the
rebrightening indicates there was a huge impulsive energy injection
$\Delta E \sim 1.8 E_0$ at $\sim 1$ hr where $E_0$ is the blast
wave energy right before the rebrightening. A typical p=2.3 value is assumed for the 
power-law distribution of electrons energies. Consistently with observations 
(Figure~\ref{fig:LC}), the X-ray flux is expected to increase by a factor of 3.
A rather smooth transition
from one solution to the other at most wavelengths is expected. However, this
is not the case at frequencies for which there is significant emission
from the reverse shock (Kumar \& Piran 2000; Kobayashi \& Sari 2000).
The sharp rise of the rebrightening implies that the typical frequency
of the reverse shock is close to the optical band. The small wiggles
before or after the major rebrightening can be explained by additional
minor energy injection or density clumps in the ambient medium.

The energy injection is probably due to a decrease in the Lorentz
factor of the outflow toward the end of the prompt GRB. This slow
outflow collides with a blast wave when the blast wave have slowed
down as  a result of sweeping up the ambient material. Energy injection
could also be caused by a long lasting activity of the central
engine (Burrows et al. 2005). The energy in the blast wave at late times ($>1h$) is larger
than that at the deceleration time by more than a factor of $\sim 2.8$.
This requires the prompt gamma-ray emission process (internal shocks) to
be more efficient than previously considered~\citep{Ioka05,Nousek06,Zhang06,Granot06}.

\subsection{Jet Break?}
The optical light curve of GRB 060206 steepened from $\alpha=0.95\pm0.02$ 
to $1.79\pm 0.11$ around $t = (3 \sim 9)\times10^4$~s. Although
this steepening resembles a jet break, the X-ray light curve from
$t=5\times10^3$ s to $7\times10^5$ s remains  consistent with a single
unbroken power law decay with $\alpha=1.35\pm0.15$ (Morris et al
2006). This clearly contradicts the monochromatic break predication by
the jet model. In the wind model $\rho(R)\propto R^{-s}$ \citep{Chevalier99}, the cooling
frequency $\nu_c$ increases as a function of time $\nu_c \propto
t^{(3s-4)/(8-2s)}$. The passage of the cooling frequency can produce
a steepening  of $\delta \alpha=(3s-4)/(16-4s)$ in the optical band
first and in the X-ray later. The observed steepening
$\delta \alpha=0.84$ gives $s\sim 2.7$. The cooling frequency rapidly
increases as $t^{1.7}$, and it will cross the X-ray band around
$\sim 10^6$ s. The observed unbroken power
law X-ray  light curve up to $t\sim 7\times10^5$ s is consistent. However,
this scenario does not account for the difference of the decay indices
between the pre-break optical ($\alpha_{opt}=0.95$) and the X-ray
($\alpha_x=1.35$) light curves, which should be on the same spectral
segment and the decay rates should be identical to be consistent with
the scenario.

A possible solution is that a blast wave initially propagates into a
constant density medium $\rho=m_p n$, and around $t\sim 3\times10^4$ s
it breaks out in a wind-type $\rho \propto R^{-s}$ medium. Using the
isotropic gamma-ray energy about $6\times 10^{52}$ergs (Palmer et
al. 2006), the transition radius is estimated in equation (5) in which
$\zeta=(1+z)/5$ and $n_1=n/10$ protons cm$^{-3}$:

\begin{equation}
R_{tr}(cm) \sim3\times10^{17} \zeta^{-1/4} (\frac{t}{3\times10^4s})^{1/4} (\frac{E}{6\times10^{52}~erg})^{1/4} n_1^{-1/4} 
\end{equation}

Although a large amount of energy is injected to the
fireball after the prompt emission, the energy dependence of $R_{tr}$
is rather weak. In this scenario, the order of the break and
observational frequencies remains as $\nu_m < \nu_{opt} < \nu_c <
\nu_x$ after the major re-brightening $t > 1$ hr. During the constant
medium phase, the time and frequency dependences of the optical and
X-ray flux are given by Equations (6) and (7) respectively (Sari, Piran
\& Narayan 1998):

\begin{eqnarray}
F_{opt}&\propto&~t^{-3(p-1)/4}\nu^{-(p-1)/2}\\
F_x&\propto&~t^{-(3p-2)/4}\nu^{-p/2}
\end{eqnarray}

With the standard value of $p=2.3$ these give
reasonable fits to the observational decay indices, while the
estimated optical spectral index $\beta \sim 0.7$ is slightly
shallower than the observations $\beta \sim 0.93\pm 0.02$. Moreover, 
a spectral break is not directly inferred between optical and X-ray
that seem to lie on a single slope.
However, as explained in the previous Section, the predicted 
$\beta \sim 0.7$ value is not significantly ruled out when a broken 
power-law is invoked.
When the
blast wave propagates into a wind medium, as the ambient density
decreases, the optical flux drops as $F_{opt}\propto t^{-\alpha_w}$
where 
\begin{equation}
\alpha_w=\frac{\bra{2s+3(p-1)(4-s)}}{(16-4s)} \sim 1.8
\end{equation}
for p=2.3 and s=2.5.
Since the X-ray band is above the cooling frequency, its
flux does not depend on the ambient density, and the X-ray decay index
does not change before and after the break-out. Considering the
scaling between the observer time and blast wave radius $t\propto
R^{(4-s)}$, the steep segment of the optical light curve requires 
that the ambient density should behave as $R^{-2.5}$ between 
$R_{tr}$ and $\sim 3 R_{tr}$.
The steepening from $\alpha=0.95$ to $1.79$ in the optical light curve
looks like a typical jet break. 
The X-ray light-curve, however, shows no sign of a change in decay slope.
This conflicts with the jet break potential interpretation based on the 
optical light-curve.
In most previous events, X-ray data are not available
at the ``jet breaks'', and the breaks are interpreted based on optical
data. The current estimate on the jet opening angle distribution might
be biased by events like GRB 060206. Consequently, the
GRB explosion energy distribution and the event rates may need to be reviewed.

\section{Conclusions}
\label{sec:conc}
We presented a comprehensive study of the high redshift GRB~060206 based on early robotic 
observations, later published photometry, GCN circulars and Swift public data.
The composite monochromatic light-curve is exceptionally well sampled and extremely structured. 
An early, $\Delta t<<t$ rebrightening suggests internal engine activity (internal shocks) continuing well beyond 
the gamma prompt emission. Later on, a high-significance {\em SED} evolution is inferred 
from multi--color $r'i'z'$ photometry carried out by the Liverpool Telescope; this is probably
associated with another pre-bump rebrightening.
One hour after the GRB a significant, steep double-peaked
rebrightening occurred, increasing the flux by a factor $\sim4$ with respect to the pre-bump values.
We interpret the cause of this rebrightening to be due to a late-time
energy injection. The energy in the blast wave at late time
($t>1h$) is a factor of 2.8 larger than that at the deceleration
time, implying a higher internal shock efficiency than previously considered.
Following the big bump, a rather typical spectral/temporal afterglow
behaviour is observed, with a possible density variation feature 
superimposed on the smooth power-law decay. We interpret the fact that
the steepening in the optical light-curve observed at $t\approx0.6d$
post-burst is not matched by similar behaviour in the X-ray
 as due to a transition in the density distribution of the ambient
medium, likely from a constant density medium to a
wind medium. In the absence of X-ray data the observed optical steepening would have been
interpreted as a typical jet-break. Since most of the so far observed
jet-breaks lack corresponding X-ray confirmation, we suggest that current
estimate on the jet opening angle distribution might be biased by
events like GRB~060206. The GRB explosion energy distribution and the
event rates are thus potentially affected.

\acknowledgments

We thank P. O'Brien, J. Osborne, B. Zhang and D. Burrows for comments
on the XRT observations and we are grateful to Prof. K.Z. Stanek for
providing the MDM photometry table.  CG and AG acknowledge their
Marie Curie Fellowships from the European Commission.  CGM
acknowledges financial support from the Royal Society.  AM
acknowledges financial support from INFN and Provincia Autonoma di
Trento.  The Liverpool Telescope is operated on the island of La Palma
by Liverpool John Moores University at the Observatorio del Roque de
los Muchachos of the Instituto de Astrofisica de Canarias.

\clearpage


\clearpage

\clearpage

\begin{deluxetable}{ccccc}     
\tabletypesize{\scriptsize}
\tablecaption{Liverpool Telescope GRB~060206 Optical Photometry\label{tab:obs}}
\tablewidth{0pt}
\tablehead{
     \colhead{Filter} & \colhead{Epoch}\tablenotemark{a} & \colhead{Exposure} & \colhead{Magnitude} & \colhead{Flux}\\
                &    \colhead{(s)}   & \colhead{(s)} &   &  \colhead{(mJy)}\\
}
\startdata
$r'$ &   319& 10  & $16.93\pm0.02$ & $0.614\pm0.011$\\
$r'$ &   341& 10  & $17.02\pm0.02$ & $0.565\pm0.010$\\
$r'$ &   363& 10  & $17.04\pm0.03$ & $0.555\pm0.015$\\
$r'$ &   415& 10  & $17.12\pm0.03$ & $0.515\pm0.014$\\
$r'$ &   437& 10  & $16.95\pm0.02$ & $0.603\pm0.011$\\
$r'$ &   459& 10  & $16.96\pm0.04$ & $0.597\pm0.021$\\
$r'$ &   574& 30  & $17.12\pm0.02$ & $0.515\pm0.009$\\
$i'$ &   785& 120 & $17.04\pm0.04$ & $0.555\pm0.020$\\
$z'$ &   885& 120 & $17.03\pm0.03$ & $0.560\pm0.015$\\
$r'$ &  1036& 120 & $17.80\pm0.02$ & $0.275\pm0.005$\\
$r'$ &  1169& 120 & $17.93\pm0.03$ & $0.244\pm0.007$\\
$i'$ &  1336& 120 & $17.23\pm0.03$ & $0.466\pm0.013$\\
$i'$ &  1487& 120 & $17.24\pm0.05$ & $0.461\pm0.021$\\
$z'$ &  1762& 120 & $17.06\pm0.03$ & $0.545\pm0.015$\\
$r'$ &  1921& 120 & $17.98\pm0.03$ & $0.233\pm0.007$\\
$r'$ &  2052& 120 & $18.02\pm0.03$ & $0.225\pm0.006$\\
$r'$ &  2185& 120 & $18.02\pm0.03$ & $0.225\pm0.006$\\
$i'$ &  2355& 120 & $17.31\pm0.04$ & $0.433\pm0.016$\\
$i'$ &  2487& 120 & $17.29\pm0.04$ & $0.441\pm0.016$\\
$i'$ &  2619& 120 & $17.31\pm0.03$ & $0.433\pm0.012$\\
$z'$ &  2807& 120 & $17.02\pm0.04$ & $0.565\pm0.020$\\
$z'$ &  2939& 120 & $16.85\pm0.04$ & $0.661\pm0.024$\\
$z'$ &  3071& 120 & $16.41\pm0.04$ & $0.991\pm0.036$\\
$r'$ &  3273& 120 & $17.04\pm0.04$ & $0.555\pm0.020$\\
$r'$ &  3405& 120 & $16.82\pm0.02$ & $0.679\pm0.012$\\
$r'$ &  3538& 120 & $16.73\pm0.02$ & $0.738\pm0.013$\\
$r'$ &  3670& 120 & $16.70\pm0.02$ & $0.759\pm0.014$\\

\enddata
\tablenotetext{a}{Time delay with respect to the GRB trigger time, $t_0=0.09302$~UT.}.
\end{deluxetable}

\clearpage

\begin{deluxetable}{llllll}     
\tabletypesize{\scriptsize}
\tablecaption{GRB~060206 SDSS-R lightcurve fit results\tablenotemark{a}\tablenotemark{b}\tablenotemark{c}\label{tab:fittime}}
\tablewidth{0pt}
\tablehead{
\colhead{Segment}&\colhead{$F_j$}&\colhead{$\alpha_1$}&\colhead{$\alpha_2$}&\colhead{$t_j$}&\colhead{Comment} \\
\\
     &    \colhead{(mJy)}   &  &   &  \colhead{(s)} &  \\
}

\startdata
 0 & n.d. & n.d. & $0.94\pm0.17$ & 0 & $\alpha_{2}$=early slope\\
 1 & $0.160\pm0.016$ & $-30$ (fixed) & $1.2\pm0.5$ & $439\pm3$ &\\
 2 & $0.076\pm0.007$ & $-30$ (fixed) & $1.0$ (fixed) & $1500$ (fixed) & \\
 3 & $0.566\pm0.020$ & $-8.0\pm0.5$  & $4.3\pm0.6$ & $3529\pm26$ & big bump\\
 4 & $0.356\pm0.027$ & $-4.9\pm0.7$  & $0.95\pm0.02$ & $5234\pm79$ & $\alpha_{2}$=post-bump slope\\
 5 & $0.029\pm0.002$ & $-3.7\pm0.7$  & $3.4\pm0.5$ & $16037\pm397$ & density bump\\
 6 & n.d. & n.d.  & $1.79\pm0.11$ & $\approx 53000$ & $\alpha_{2}$= slope after final break\\
\enddata
\tablenotetext{a}{Refer to the text following equation 1 for an explanation.}
\tablenotetext{b}{The $n$ exponent is fixed to 2.5 for all the segments.}
\tablenotetext{c}{The segment 6 has been fitted separately with a simple power-law.}

\end{deluxetable}

\begin{deluxetable}{lllll}     
\tabletypesize{\scriptsize}
\tablecaption{Fitted spectral slopes from multi--epoch SED (Figure \ref{fig:SED})\label{tab:spectra} and text}
\tablewidth{0pt}
\tablehead{
\colhead{$Time (s)$} & \colhead{$\beta_O$} & \colhead{$\beta_X$} & \colhead{$\beta_{O-X}$} & \colhead{Comments}\\
 & & & & \\
}

\startdata
$785$ & $1.42\pm0.58$\tablenotemark{a} & $1.15\pm0.55$\tablenotemark{b} & $0.81\pm0.03$ & $\chi_{O-X}^{2}/d.o.f=4.26/4$ \\
$\approx1800$ & $2.05$\tablenotemark{c} & $1.15\pm0.55$\tablenotemark{b} & $n.d.$ & $\beta_{O}\neq\beta_{X}$ \\
$5680$ & $0.84\pm0.14$\tablenotemark{d} & $0.84\pm0.15$ & $0.93\pm0.02$ & $\chi_{O-X}^{2}/d.o.f=3.46/10$\\
$\approx108000$ & $0.7\pm0.3$\tablenotemark{e} & $n.c.$ & $n.c.$ & $\chi_{IR-O}^{2}/d.o.f=0.85/1$  \\
\enddata
\tablenotetext{a}{Excluding B (UVOT) value affected by {\em HI} absorption}
\tablenotetext{b}{Assumed from{\em XRT} spectrum in $t=141s-341s$. See text}
\tablenotetext{c}{Calculated from $r'i'z'$ points interpolated according the additional bump hypothesis (model-dependent)}
\tablenotetext{d}{From back-extrapolations in $JHK$ (PAIRITIEL) and $R$ (MDM)}
\tablenotetext{e}{SUBARU-IR \& MDM-R}

\end{deluxetable}

\clearpage






\end{document}